\documentclass[doublecol]{epl2}
\usepackage{graphicx,color,amsmath,bm,subfigure,amsmath,amssymb,verbatim,MnSymbol}

\newcommand{\D}{\mathrm{d}}
\newcommand{\e}{\mathrm{e}}

\newcommand{\be}{\begin{equation}}
\newcommand{\ee}{\end{equation}}
\newcommand{\bea}{\begin{eqnarray}}
\newcommand{\eea}{\end{eqnarray}}

\newcommand{\kbt}{k_{\mathrm{B}}T}

\begin{document}


\title{Complex Fluids with Mobile Charge-Regulating Macro-Ions}
\shorttitle{Charge regulation of mobile macro-ions}
\author{Tomer Markovich \inst{1,2} \and David Andelman \inst{1} \and Rudi Podgornik \inst{1,3,4,5}}
\shortauthor{T. Markovich \etal}

\institute{
\inst{1} Raymond and Beverly Sackler School of Physics and Astronomy, Tel Aviv University, Ramat Aviv 69978, Tel Aviv, Israel \\
\inst{2} DAMTP, Centre for Mathematical Sciences, University of Cambridge, Cambridge CB3 0WA, United Kingdom \\
\inst{3} Department of Theoretical Physics, J. Stefan Institute, 1000 Ljubljana, Slovenia \\
\inst{4} Department of Physics, Faculty of Mathematics and Physics, University of Ljubljana, 1000 Ljubljana, Slovenia\\
\inst{5} School of Physical Sciences, UCAS, University of Chinese Academy of Sciences, Beijing 100049, China}


\abstract{
We generalize the concept of charge regulation of ionic solutions, and apply it to complex fluids with mobile macro-ions having internal non-electrostatic degrees of freedom.
The suggested framework provides a convenient tool for investigating systems where
mobile macro-ions can self-regulate their charge ({\it e.g.}, proteins).
We show that even within a simplified charge-regulation model, the charge dissociation equilibrium results in different and notable properties.
Consequences of the charge regulation include a positional dependence of the effective charge of the macro-ions,
a non-monotonic dependence of the effective Debye screening length on the concentration of the monovalent salt,
a modification of the electric double-layer structure, and buffering by the macro-ions of the background electrolyte.}

\maketitle

\section{Introduction}

{\it Charge regulation} (CR) is a mechanism introduced in the 1970's~\cite{NP-regulation} to explain charged colloidal systems, and is related to association/dissociation of mobile ions to/from macromolecular surfaces modelled as  external bounding surfaces. CR leads to a rather complex self-consistent relation between the electrostatic surface-potential and surface-charge density. This relation depends on the salt concentration, pH, system geometry and other details of the association/dissociation process~\cite{Borkovec1,Borkovec2,CR1,Safynia}. For example, we have recently shown~\cite{CR1} that the CR boundary condition is inherently a separate boundary condition
as compared with the constant potential and constant charge ones.
It yields a disjoining pressure between two charge surfaces with a unique scaling, which does not reduce to the constant charge or constant potential cases.

In particular, it leads to different scaling behavior of the osmotic pressure.
The CR can be modelled either by the chemical association/dissociation equilibrium of surface binding sites (law of mass action)~\cite{Regulation2,Regulation3,Regulation4,Regulation5}, or equivalently, via a surface free-energy approach~\cite{Olvera,Olvera2,Natasa3,epl,diamant,maarten,everts}.

The CR process is widely recognized to be an essential process influencing the stability of electrostatic double-layers, inter-surface electrostatic forces~\cite{instab,Harries}, and the dissociation of amino acids and related protein-protein interactions~\cite{Leckband,Lund,Fernando}. In addition, the CR affects the stability of proteinaceous viral shells~\cite{Nap}, adsorption of polyelectrolyte and the behavior of polyelectrolyte brushes~\cite{Netz-CR,Borukhov,Kilbey,Zhulina,maarten2008}, as well as the interactions between charged membranes~\cite{membranes1, membranes2, membranes3}.

\begin{figure}[h!]
\centerline{\includegraphics[scale=0.28]{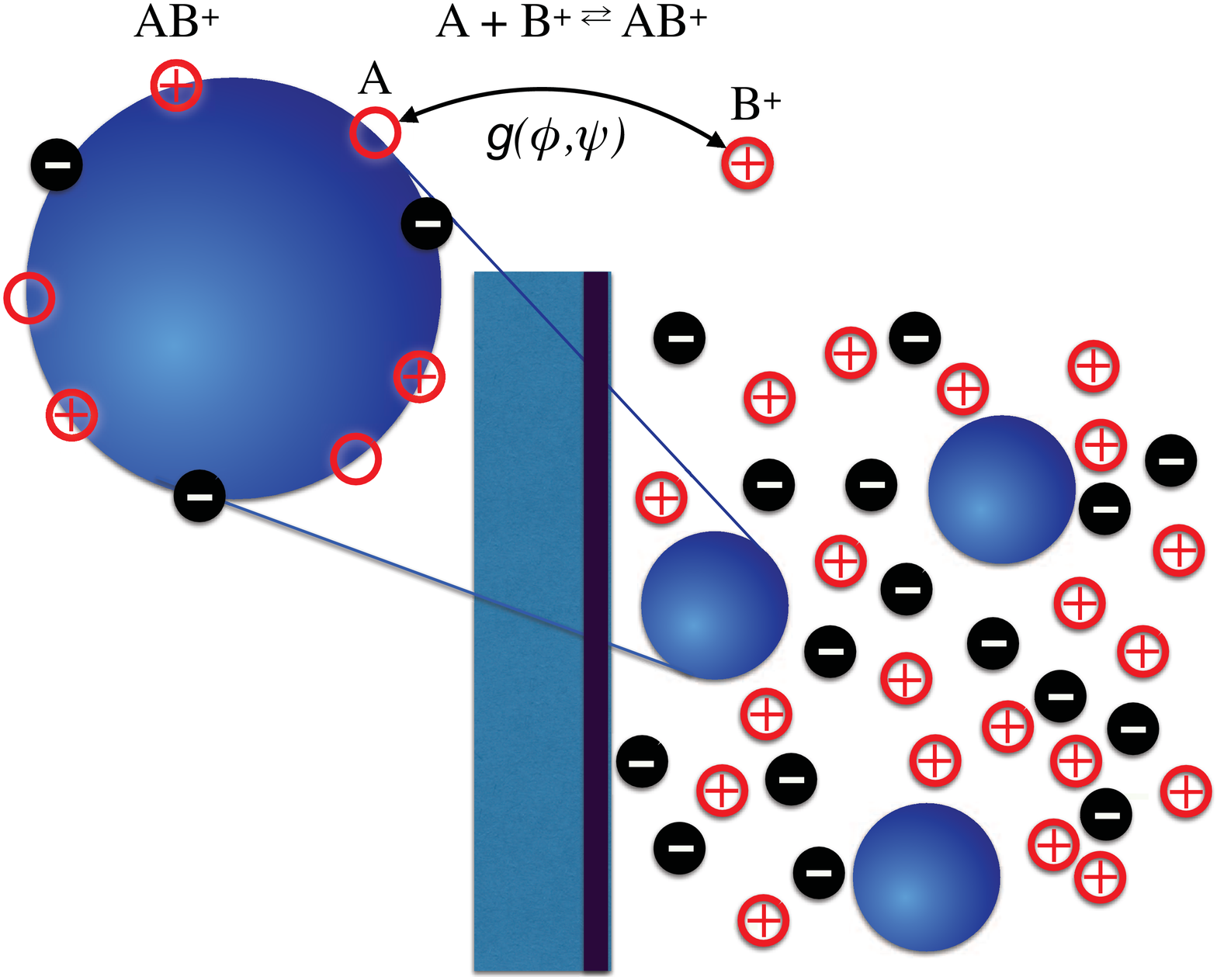}}
\caption{\textsf{(color online)~Schematic setup of an aqueous solution composed of small monovalent ions (depicted red for cations and black for anions), and mobile macro-ions (blue) with dissociable  groups (shown under magnification) undergoing a dissociation reaction $\rm A + B^{+} \rightleftharpoons AB^{+}$.  The black anions are either dispersed in solution or bound to the macro-ion, while the red cations are always free. In the magnified view, the empty red circles represent the dissociable/adsorbing sites on the blue macro-ion. Furthermore, the macro-ions are described by an internal free energy, $g(\phi,\psi)$, that depends on the annealed fraction of charge groups on the macro-ion, $\phi$, and the local electrostatic potential, $\psi$.}
\label{Fig1}}
\end{figure}

In this Letter, we focus on the interplay between charge regulation of {\it mobile} macro-ions and the corresponding screening properties in a bathing solution of monovalent salt, see Fig.~\ref{Fig1}. This differs from the standard framework of CR processes, where the dissociable moieties are confined to fixed bounding macromolecular surfaces, while the intervening ionic solution is assumed to be fully dissociated. Attempts in this direction were only made so far within the cell model approximation~\cite{boon,maarten2006},
but these do not fully capture the collective effects of the macro-ions.

Our formulation allows a straightforward generalization of the conceptual framework, as applicable to complex fluids composed of annealed colloidal particles dispersed in electrolyte solutions. The new framework broadly shows anomalous screening, significantly modifies  the double layer structure and can exhibit buffering, {\it i.e.}, the macro-ion charge adjusts in such a way that the behavior of the system becomes independent of the salt concentration. Among others, our results can have important consequences on the structure and function of nano-particles or proteins in enzyme nano-reactors in situations when their charge groups are dissociable.

\section{Model}
Consider an ionic aqueous solution composed of a 1:1 monovalent salt with bulk concentration $n_b$
and mobile, originally neutral, CR macro-ions of bulk concentration $p_b$.
One can consider several charging models of the macro-ions.
These models describe a CR process that is governed by an additional free-energy contribution, taking into account the internal degrees of freedom of the macro-ion.
Once the macro-ions are immersed in the solution they may adsorb or release small ions, thus changing the amount of ``free" ions in the solution.

On a mean-field level, the grand-canonical free energy can be written as a sum of the electrostatic energy of the three ionic species in solution (monovalent cations/anions and macro-ions) and their entropy of mixing~\cite{Safynia} assuming a dilute 
solution,
\begin{eqnarray}
\label{g2}
\nonumber F &=&  \int_{V}\D^3r\Bigg( -\frac{\varepsilon}{8\pi}\left( \nabla\psi\right)^2 +   \sum_{i=\pm}e_in_i \psi \,+ \,p\,g\!\left(\phi,\psi\right)  \nonumber\\
&+& \kbt~\sum_{i=\pm} \Big[ n_i\ln\left(n_i a^3\right) -n_i \Big] - \sum_{i=\pm} \mu_i n_i   \nonumber\\
&+& \kbt ~\Big[ p\ln\left(p~\!a_p^3\right) -p \Big] -  \mu_p p  \Bigg) \, ,
\end{eqnarray}
where $\varepsilon$ is the solution dielectric constant, $k_BT$ is the thermal energy, $\psi({\bf r})$, $n_\pm({\bf r})$ and $p({\bf r})$ are, respectively,  the local mean-field electrostatic potential, and the number concentration (per unit volume) of the three charge species. The monovalent cations and anions are characterized by their charge $e_\pm=\pm e$ ($e$ is the unit electron charge), size $a$ and chemical potential  $\mu_\pm$. The subscript $p$ stands for the macro-ion degrees of freedom, so that the macro-ion size (radius) is $a_p$ and their chemical potential is $\mu_p$. For simplicity, we set $a_p = a$, where $a$ is the single microscopic length scale entering the model.

The only non-standard term in Eq.~(\ref{g2}), extending in a significant way previous approaches, is the $pg(\phi,\psi)$ term.
It depends explicitly on the free energy of a single macro-ion, $g(\phi,\psi)$,
accounting for its internal degrees of freedom and evaluated at position $\bf r$ of the macro-ion.

This free-energy approach describes the collective, average CR response of the macro-ions.
Therefore, it depends also on their local spatially-dependent concentration.
In this sense, our collective description of the CR macro-ions differs significantly from other approaches based on the cell model approximation~\cite{maarten2006,boon}, as these latter approaches do not fully capture the collective effect of the macro-ions.

In our CR model as depicted in Fig.~\ref{Fig1}, each macro-ion has a fixed number, $M_p$, of charge groups,
contributing a fixed negative charge $-e M_p$.
In addition, there are $N_p$ sites that can be either dissociated (neutral) or associated
(positively charged by adsorbing a positive monovalent ion from the solution).
In its initial state, before the macro-ion is immersed in the solution, $M_p$ of its CR sites are filled with positive ions.
After the macro-ion is inserted in the solution, its effective charge on the $N_p$ sites depends on the annealed dissociation fraction $0\le \phi \le 1$, yielding a positive charge $e N_p \phi$. Hence, the total charge on the macro-ion, $e_p$, is given by $e_p \equiv e (N_p \phi - M_p)$.

Using the above model for the annealed charges on the macro-ions, the internal free-energy of the CR macro-ions, $g$, corresponds to the Langmuir-Davies isotherm and can be written as
\begin{eqnarray}
\label{b3}
g(\phi,\psi) &=& e_p\psi - (\alpha + \mu_+) N_p \phi ~ \nonumber\\
&+&  k_B T N_p \, \Big( \phi\ln \phi + \left(1-\phi \right)\ln\left(1-\phi \right)\Big).
\end{eqnarray}
The above introduced $\alpha$ parameter can be regarded as the chemical potential change for the dissociation process $$\rm A + B^{+} \rightleftharpoons AB^{+}\, ,$$ where $\rm AB^{+}$ is the dissociable moiety and $\rm B^{+}$ is the dissociated monovalent cation (see Fig.~\ref{Fig1}).
\footnote{This CR model is nothing but two uncorrelated CR processes,
$\rm A_1 + B_1^{+} \rightleftharpoons A_1B_1^{+}$ and $\rm A_2 + B_2^{-} \rightleftharpoons A_2B_2^{-}$, with very strong anion adsorption
(see Eq.~(19) of Ref.~\cite{CR1}).}
Note that within our convention, $\alpha>0$ ($\alpha<0$) corresponds to an attraction (repulsion) between cations and the macro-ion.  Furthermore, the electrostatic part of this free energy, $e_p\psi $, contains the electrostatic energy of the fixed charges, $- e M_p \psi$, as well as that of the dissociable ones, $(e N_p \phi) \psi$.

We further assume that $N_p=2M_p$, implying that the charge on the  macro-ion varies symmetrically around zero charge,  $e_p=e z_p(2\phi-1)$, with $-ez_p\le e_p\le ez_p$ and $z_p = M_p=  N_p/2$ being the maximal valency. This choice of a symmetric charge distribution on the colloid particle  is chosen for convenience, but can be easily  extended to models with different numbers of dissociation sites and/or fixed charges~\cite{Natasa3}.

Taking the variation of the free energy, Eq.~(\ref{g2}), with respect to the densities, $n_{\pm}$ and $p$, we obtain the corresponding Euler-Lagrange (EL) equations describing the thermodynamic equilibrium
\begin{eqnarray}
\pm e \psi + k_BT \ln(n_{\pm} a^3) - \mu_{\pm}= 0, \nonumber\\
g(\phi,\psi) + k_BT \ln(p~\!a^3) - \mu_{p}= 0,
\label{g5}
\end{eqnarray}
while the variation with respect to $\phi$ yields an EL equation in the form
\begin{eqnarray}
k_BT \ln\left(\frac{\phi}{1-\phi}\right) + e\psi - \alpha - \mu_+ = 0,
\label{EL3}
\end{eqnarray}
which is the standard {Langmuir-Davies} isotherm for the dissociation degree $\phi$. Finally, the EL equation for $\psi$ results in a modified Poisson-Boltzmann (PB) equation of the form
\begin{eqnarray}
\label{g6}
-\frac{\varepsilon}{4\pi}\nabla^2\psi &=& \!\!\!\! e\left( n_{+} - n_{-} \right) + p\, \frac{\partial g(\phi,\psi)}{\partial \psi} \\
\nonumber&=& \!\!\!\! e \Big[ n_b^{+} \e^{-\beta e\psi} - n_b^{-}\e^{\beta e\psi} \\
\nonumber&-& \!\!\!\! z_p p_b \e^{\beta z_p e\psi}  \left( \frac{1 + \e^{-2\Lambda(\psi)}}{1 + \e^{-2\Lambda(0)}}\right)^{2z_p}
\!\!\!\!\!\!\tanh{\Lambda(\psi)}\Big] \, ,
\end{eqnarray}
where $\beta=1/(k_B T)$ and $\Lambda(\psi) \equiv \beta \left[ e \psi - ({\mu}_{+} + {\alpha})\right]/2$.
The right-hand side of the above equation is equal to the total charge density,
$\rho(\psi)$, and the second equality is obtained by using Eqs.~(\ref{g5}) and (\ref{EL3}).
We have also defined the bulk densities of the cations/anions and macro-ions as $n_b^{\pm} \equiv \exp(\beta\mu_\pm)/a^3$ and
$p_b \equiv \exp(\beta [\mu_p - g(\psi=0)])/a^3$, respectively.
We remark that electroneutrality is enforced by requiring that the integral over all charges (mobile and fixed) is zero.

In the general case represented by Eq.~(\ref{g6}), the charge of the macro-ion, $e_p$,
can be written as
\begin{equation}
e_p({\bf r})=\frac{\partial g(\phi,\psi)}{\partial \psi}
\label{g6_1} ,
\end{equation}
where the charge $e_p$ depends explicitly on $\phi$ and $\psi$ and implicitly on the position ${\bf r}$. This charge can either be negative or positive, varying in the range  $-e z_p \le e_p \le e z_p$.
We note that Eq.~(\ref{g6}) reduces to the standard PB equation in the limit where the free energy $g$ has only a purely electrostatic origin, which depends on $\psi$ but is independent on $\phi$. This is written as $g(\psi) = e_p \psi$, where $e_p=e z_p$ is the effective macro-ion charge and $z_p$ is its valency. Note that the charge density depends only on the product $e_p\,p$.

\begin{figure}[t!]
\centerline{\includegraphics[width=0.5\textwidth]{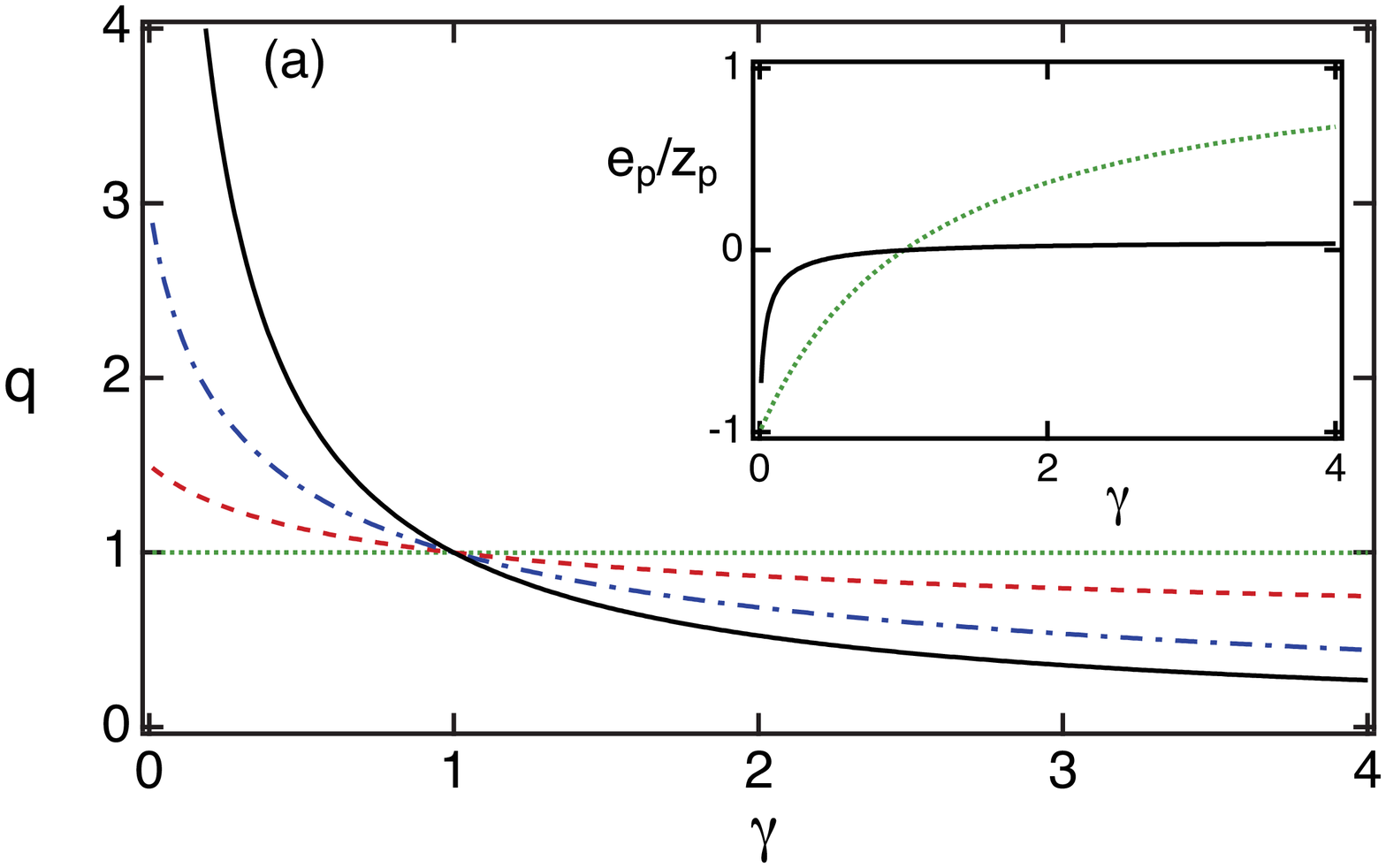}}
\centerline{\includegraphics[width=0.5\textwidth]{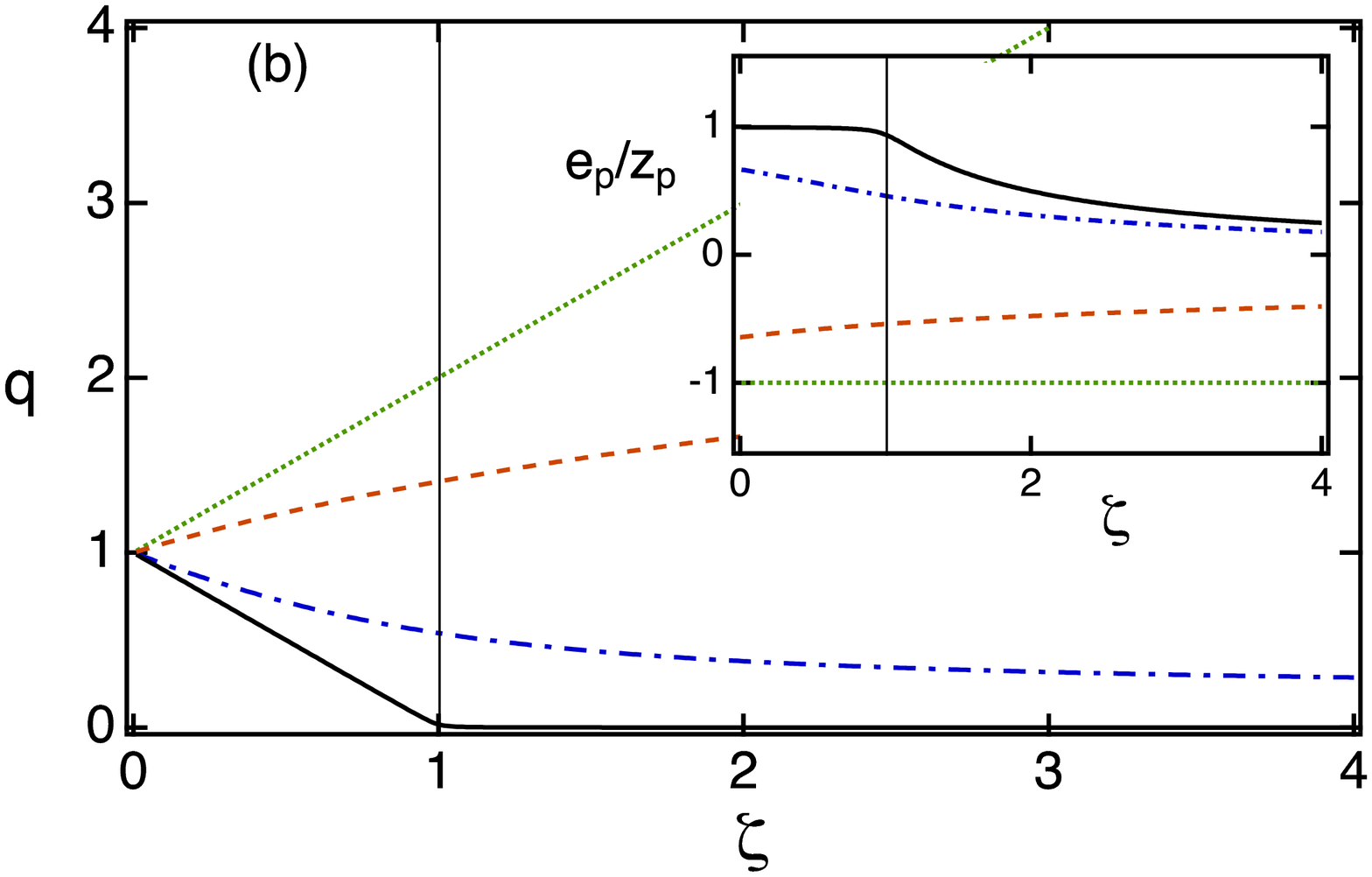}}
\caption{(color online)~ \textsf{(a)~Charging parameter of the cations $q(\zeta,\gamma) = n_b^{+}/n_b$, and (inset) the effective charge of macro-ions, $e_p = {\partial g}/{\partial \psi}\vert_{\psi = 0}$, in the bulk (in units of $z_p$), as function of the rescaled concentration of the monovalent salt, $\gamma  = n_b a^3\e^{\beta{\alpha}}$, for different values of the ratio $\zeta = {z_p p_b}/{n_b}$, with $\zeta =$ 20 (full line, black), 2 (dash dotted line, blue), 0.5 (dashed line, red), 0 (dotted line, green). In the inset, for clarity purposes, only $\zeta=20$ (full line, black) and $\zeta =0$ (dotted line, green) are shown. The latter corresponds to a simple monovalent salt without the macro-ions ($p_b=\zeta = 0$). (b)~Charging parameter $q$ and  the effective charge $e_p$ (inset) of macro-ions as function of $\zeta$ for different values of $\gamma =$ 500 (full line, black), 5 (dash dotted line, blue), 0.3 (dashed line, red), and 0 (dotted line, green), in the main figure and in the inset.  For $\gamma \gg 1$, the {\it buffering action} of the macro-ions can clearly be noticed as $q$ ceases to depend on the salt concentration.
}
\label{Fig2}}
\end{figure}

The total charge density, $\rho$, in the right-hand-side of Eq.~(\ref{g6}) is related to the pressure through the first integral of the modified PB equation, Eq.~(\ref{g6}), and can be cast as (see Ref.~\cite{Maggs} for further details)
\begin{eqnarray}
\rho(\psi) = - \frac{\partial P_0(\psi)}{\partial \psi} \, .
\label{bcskygk1}
\end{eqnarray}
Here, $P_0(\psi)$ is the MF entropic contribution to the pressure (van 't Hoff ideal pressure) of all mobile species in the solution,
given by $P_0(\psi)=k_B T\left[n_{+}(\psi) + n_{-}(\psi)  + p(\psi)\right]$, or explicitly as
\begin{equation}
P_0(\psi) =  k_B T n_b \Big[q \e^{-\beta e\psi} + \e^{\beta e\psi} +\frac{p_b}{n_b}\left( \frac{1 + \e^{-2\Lambda(\psi)}}{1 + \e^{-2\Lambda(0)}} \right)^{2z_p}\!\!\!\!\!\e^{\beta z_p  e\psi}\Big],
\label{bcskygk2}
\end{equation}
with $q \equiv n_b^{+}/n_b$ defined as the {\it charging parameter}. For systems with a planar symmetry, the pressure can be written~\cite{Dan,Safynia,Maggs} as the sum of the Maxwell stress and $P_0$,
$$P = -\frac{\varepsilon}{8\pi}\psi'^2(z) +  P_0(z).$$

\section{Charge regulation and screening}
In the bulk $\psi = 0$ with $\Lambda_b = \Lambda(\psi = 0)$, and the electroneutrality condition is written in the form
\begin{eqnarray}
\label{g9}
e\left( n_b^{+} - n_b^{-} \right) + p_b \cdot \frac{\partial g(\phi,\psi)}{\partial \psi}\Bigg|_{\psi{=}0, \phi{=}\phi_b} =~ 0.
\end{eqnarray}
The bulk number concentration of macro-ions is $p_b$ and that of the anions is defined to be $n_b^{-}=n_b$.
The number concentration of the cations, $n_b^{+}$, is obtained by solving the electroneutrality condition of Eq.~(\ref{g9}).
It is instructive to consider two sub-cases: (i) $n_b^{+}\geq n_b$ and, (ii) $n_b^{+}< n_b$.
The former (latter) corresponds for macro-ions that are positively (negatively) charged in the bulk,
and the number concentration of cations (anions) comes from two sources: an $n_b$ ($n_b^+$) amount coming from the monovalent salt,  and a contribution $n_b^{+}-n_b>0$ ($n_b-n_b^{+}>0$) originating from the dissociation process of the macro-ions.

Using the charge density from Eq.~(\ref{bcskygk1}) for $\psi{=}0$ and the charging parameter, the electroneutrality condition leads directly to a quadratic equation for $q$. Its solution yields
\begin{eqnarray}
q\left(\zeta, \gamma\right)\!\!\!\!&=&\!\!\!\!  {\textstyle\frac12} \Big( 1 - \zeta - \gamma^{-1}   \nonumber\\
&+& \sqrt{\left( 1 - \zeta - \gamma^{-1}\right)^2+ 4(1+\zeta)\gamma^{-1}} ~\Big).
\label{enc}
\end{eqnarray}
with two new parameters defined as $\zeta \equiv {z_p p_b}/{n_b}$ and $\gamma  \equiv n_b a^3\e^{\beta{\alpha}}$.  In Fig.~\ref{Fig2} we present the dependence of $q$  on $(\zeta, \gamma)$. This parametrization corresponds to setting the ratio $n_b/p_b$ fixed, while varying  the monovalent salt concentration, $n_b$. Note that both concentrations, $n_b$ and $p_b$, get rescaled by the Boltzmann factor of the dissociation parameter, $\alpha$.
In most physical systems the macro-ions concentration is a few orders of magnitude smaller than the electrolyte concentration, and qualitatively $\zeta$ ranges between $0.01$ and $10$,
while $\gamma$ has much wider range that roughly varies between $0.001$ and $1000$.

For the realistic case of $\zeta \ll 1, \gamma^{-1}$, the charging parameter is
\begin{equation}
q(\zeta, \gamma) \simeq 1 +  \frac{1-\gamma}{1+\gamma} \zeta \, .
\label{cbskugybcf1}
\end{equation}
The vanishing concentration of the macro-ions is obtained when $q = 1$, and corresponds to salt-only bulk without any macro-ions, $p_b = 0$.
In the other limit of large  macro-ion concentration, the system shows a {\it buffering capacity}, because  the rescaled cation concentration tends to a constant for { $\zeta \gg 1, \gamma^{-1}$ }
\begin{equation}
\gamma~ q(\zeta, \gamma) \longrightarrow 1 + \frac{2}{\zeta}\left( 1-\gamma^{-1}\right).
\label{cbskugybcf}
\end{equation}
To the lowest orders, it is independent of the concentration of the ionic species.

For large salt concentrations, $\gamma \gg 1$ and $\gamma \gg \zeta$,  the macro-ions function as a buffer.
For $\zeta < 1$, the charging parameter is linear in $\zeta$. Namely,  $q = 1 - \zeta$. However, for $\zeta \geq 1$, the scaling is $q \sim \gamma^{-1}$.
The buffering action of the macro-ions results in negligible values of $q$, {\it i.e.,} almost all the cations are adsorbed onto the macro-ions as $\zeta \geq 1$,  irrespective of the exact value of the high salt concentration, see Fig.~\ref{Fig2}.
In the opposite limit of $\gamma \ll 1$, the lowest-order dependence assumes the form
\begin{equation}
q(\zeta, \gamma) \longrightarrow (1+\zeta) .
\end{equation}

Different amounts of monovalent salt can effectively charge the macro-ion with either sign of the charge, exhibiting charging curves that show a pronounced asymmetry with respect to the uncharged state. The effective charge of the macro-ions in the bulk, $e_p = {\partial f}/{\partial \psi}\vert_{\psi = 0}$, trails the functional dependence of the charging parameter, but shows (inset Fig.~\ref{Fig2}(a)) only a marginal dependence on $\zeta$ as $\gamma$ is varied. In the opposite case, a substantial variation in $e_p$ is seen (inset Fig.~\ref{Fig2}(b)) as a function of $\zeta$, while $\gamma$ is held fixed.

\begin{figure}[t!]
\centerline{\includegraphics[width=0.48\textwidth]{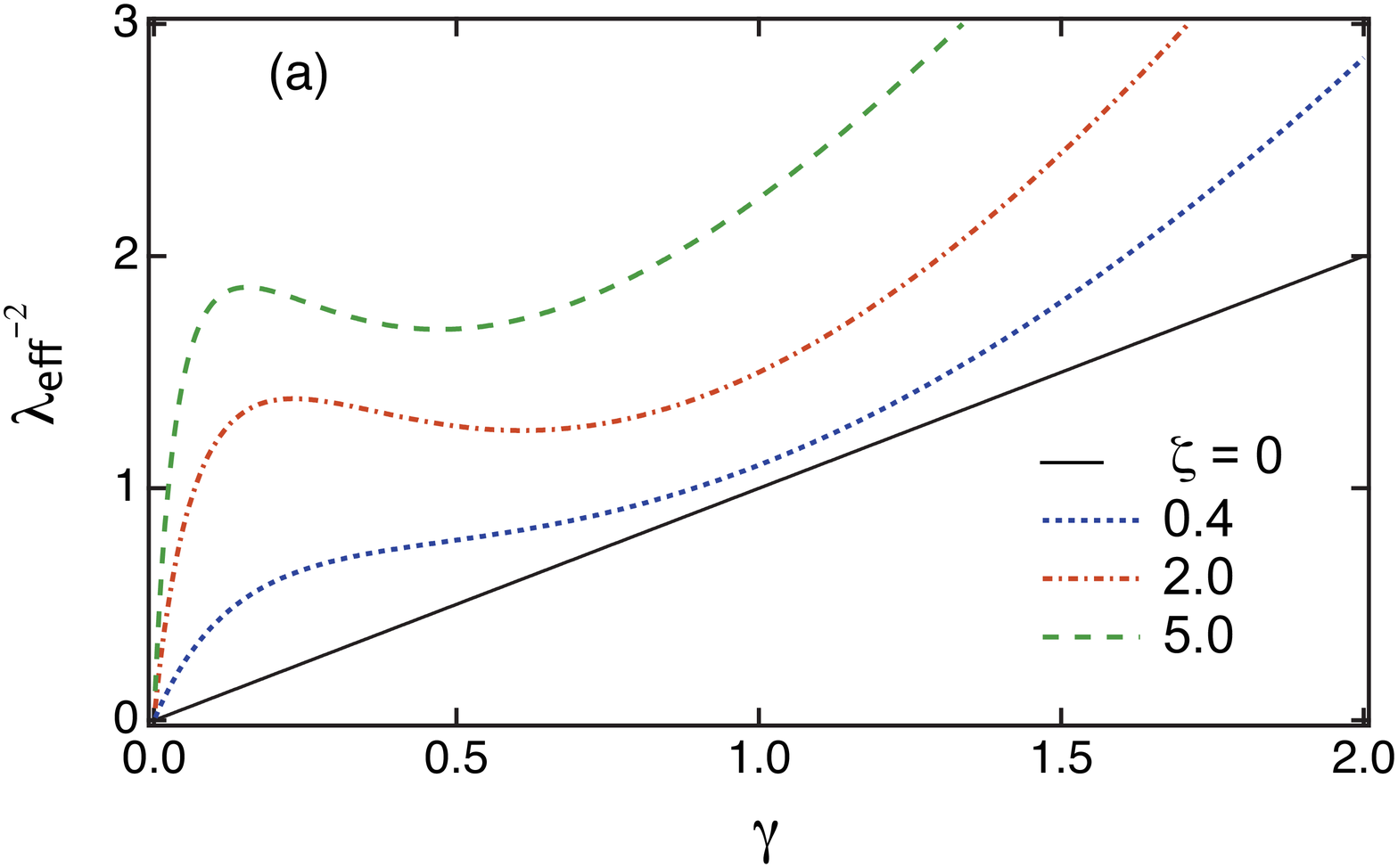}}
\centerline{\includegraphics[width=0.48\textwidth]{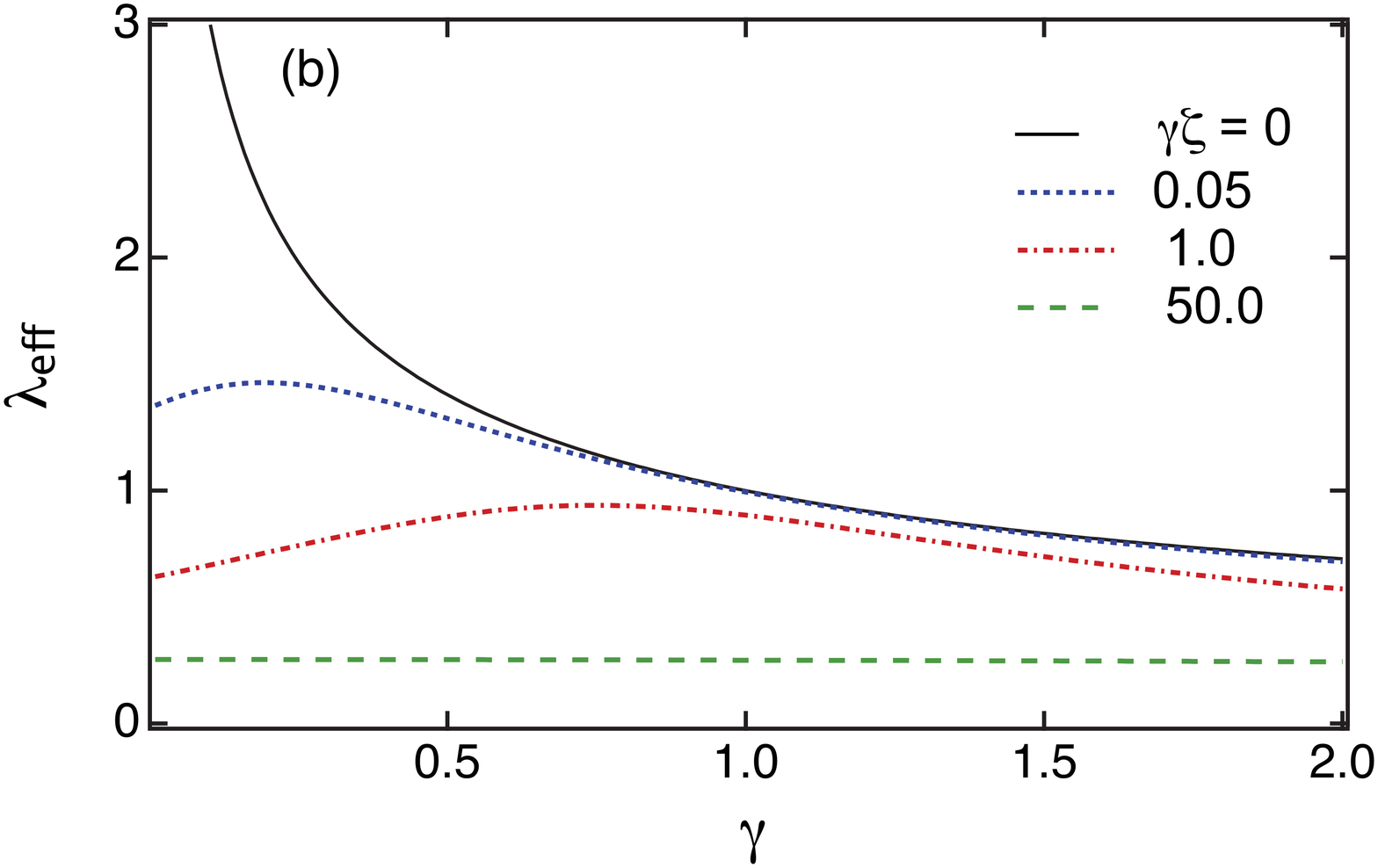}}
\caption{(color online). \textsf{(a) The effective $\lambda_{\rm eff}^{-2}$ rescaled in units of $8\pi \ell_{\rm B} a^{-3} \e^{-\beta{\alpha}}$, as a function of $\gamma$ (the rescaled salt concentration), for different values of $\zeta$ (defined as the ratio between macro-ion and salt concentrations). (b) The effective Debye length $\lambda_{\rm eff}$ rescaled in units of $({8\pi \ell_B a^{-3} \e^{-\beta{\alpha}}})^{-1/2}$, as a function of $\gamma$ but for different values of the rescaled macro-ion concentration, $\gamma\zeta= z_p p_b a^3 \e^{\beta {\alpha}}$. In both parts, the maximal macro-ion valency is set to $z_p=25$, and the plots show non-monotonic behavior as function of the salt concentration, $n_b$, while the full lines correspond to the pure salt case with $\zeta = 0$. The effective Debye length $\lambda_{\rm eff}$ is always below $\lambda_{\rm D}$, resulting in a more pronounced screening.}
\label{Fig3}}
\end{figure}

The charging equilibrium of the macro-ions can modify the Debye screening length of the electrolyte solution. Expanding the charge density $\rho(\psi)$ of Eq.~(\ref{bcskygk1}) to linear order in the electrostatic potential $\psi$, and taking into account the electroneutrality condition in the bulk where $\psi_b=0$, yields a modified Debye-H\"uckel equation with
\begin{eqnarray}
\lambda_{\rm eff}^{-2}\!\!&=&\!\!\!\!-\frac{4\pi}{\varepsilon} \frac{\partial \rho(\psi)}{\partial \psi}\Bigg|_{\psi = 0}\!\!\!\!= \nonumber\\
&& {\textstyle\frac{1}{2}} \lambda^{-2}_{\rm D} \Big( 1 + q + \frac{(1-q)^2}{\zeta} (z_p-{\textstyle\frac{1}{2}}) + {\textstyle\frac{1}{2}} \zeta \Big),
\label{kappae}
\end{eqnarray}
where $\lambda_{\rm eff}$ is the effective Debye screening length, $\lambda_ {\rm D} = (8\pi \ell_{\rm B} n_b)^{-1/2}$ is the standard Debye screening length for a monovalent salt, $\ell_{\rm B}=\beta e^2/\varepsilon$ is the Bjerrum length, and $q=q(\zeta,\gamma)$ was introduced in Eq.~(\ref{enc}). Apart from the parameters that define $q$, the effective screening length depends also on $z_p$, the macro-ion maximum valency {that in physical situations may vary between $1$ and $100$}.
Clearly, as the bulk concentration $p_b$ of the macro-ions vanishes ($q \to 1$ for $\zeta \rightarrow 0$), the effective Debye length approaches from below the standard one, $\lambda_{\rm eff} \rightarrow \lambda_{\rm D}$.
Note that the effective Debye length, which is the natural electrostatic length-scale, gives a limit of validity for the theory,
$a < \lambda_{\rm eff}$, because $\lambda_{\rm eff}$ is a bulk averaged quantity that cannot be smaller than the microscopic length-scale, $a$.

We note that for $\gamma \to 0$, where the standard Debye screening length of the salt diverges, $\lambda_{\rm D}\sim n_b^{-1/2}$, the effective screening length $\lambda_{\rm eff}$, on the other hand, converges to a finite value that depends on the rescaled concentration of the macro-ions, $\sim(\gamma\zeta)^{-1/2}$ (Fig.~\ref{Fig3}), and screening is solely due to the CR mechanism of the macro-ions.
When $\gamma\zeta$ increases, $\lambda_{\rm eff}$  reaches a maximum as a function of $\gamma$, and then decays towards the standard $\lambda_{\rm D}$.
For large $\gamma$ and for fixed $\zeta = 1$,
$\lambda_{\rm eff} \to \lambda_{\rm D}/\sqrt{( 1 + z_p)/2}$, which is equivalent to  screening for non-symmetric $1:z_p$ electrolytes. Notably, the macro-ions enhance the screening effect, leading to a screening length, $\lambda_{\rm eff}$, that is always below the corresponding Debye length, $\lambda_{\rm D}$, as is shown in Fig.~\ref{Fig3}.

\begin{figure}[t!]
\centerline{\includegraphics[width=0.48\textwidth]{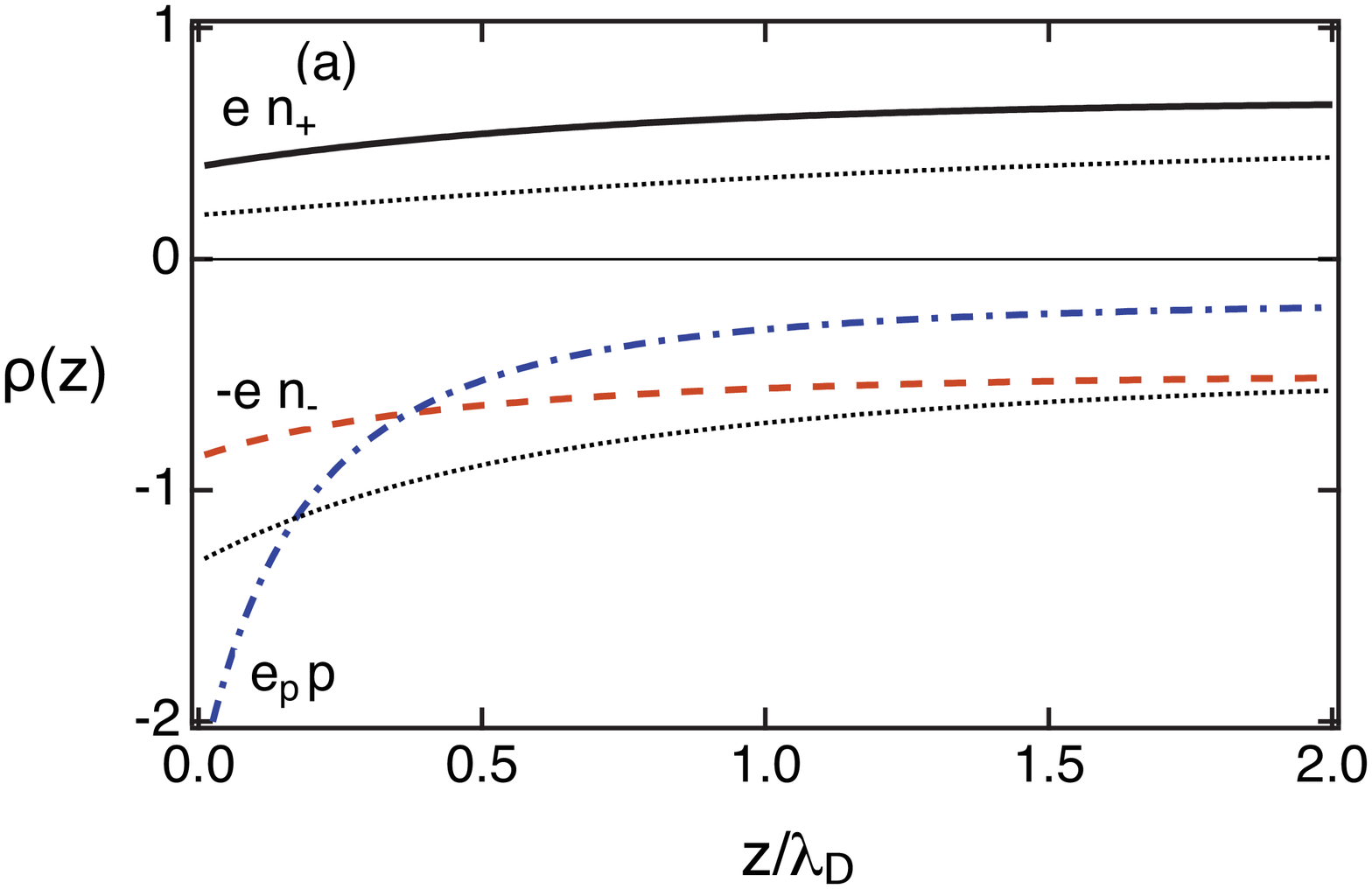}}
\centerline{\includegraphics[width=0.48\textwidth]{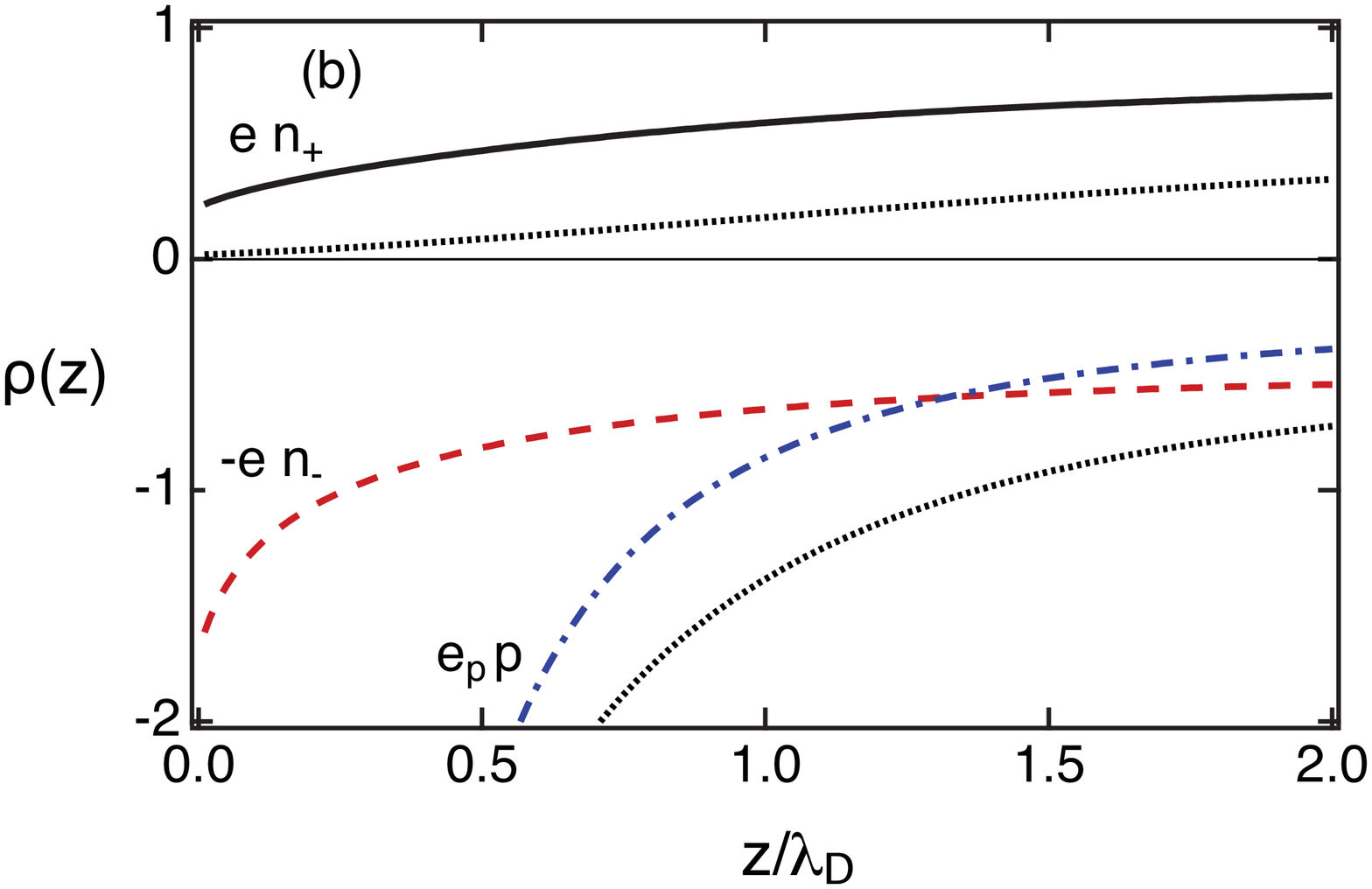}}
\caption{(color online). \textsf{(a) The three contributions to the total charge density, $\rho(z)=en_{+} - en_{-}+e_p p$, in units of $e$ and as function of the rescaled distance from the surface, $z/\lambda_{\rm D}$: cation charge density $en_{+}$ (full line, black), anion charge density, $-en_{-}$ (dashed line, red), and the macro-ion charge density, $e_p\, p$ (dash-dotted line, blue), for $\zeta = 2$, $\gamma = 0.5$, $z_p = 10$ and $\Xi = \lambda_{\rm D}/ \ell_{\rm GC}=1$. The dotted lines correspond to the cation and anion densities for the salt-only case (no macro-ions, $p_b=0$). (b) Same as (a) except that the concentration ratio was increased to $\zeta = 4$ and $\Xi=5$. For larger $\zeta$ values, most of the positive surface charges are screened by the macro-ions in solution, while the monovalent ion concentration varies less than for the salt-only case ($p_b=0$, represented by the dotted lines).
}
\label{Fig4}}
\end{figure}

\section{Charge regulation and the structure of the electric double layer}
While the screening length  quantifies the mobile charge  distribution far from external charge sources, the vicinal partitioning of the ions needs to be calculated from the full PB equation, Eq.~(\ref{g6}).  As a typical example, we assume a charged planar wall located at $z=0$, that carries a fixed surface-charge density $\sigma>0$, so that all the quantities depend only on the normal coordinate $z$ (see Fig.~\ref{Fig1}).

The first integral of the extended PB equation, Eq.~(\ref{g6}) yields the Grahame equation~\cite{Dan,Safynia}, with the surface potential $\psi(z{=}0) = \psi_s$ and its derivative $\beta e \psi'(z{=}0) = -2/\ell_{\rm GC}$, where $\ell_{\rm GC} = e/(2 \pi \ell_{\rm B} |\sigma|)$  is the Gouy-Chapman length, and $\ell_{\rm B} = \beta e^2/\epsilon$ has been defined above as the Bjerrum length. Then, an explicit solution of the PB equation is obtained by performing a second integration, yielding the relation, $z=z(\psi)$,
\begin{equation}
z = \sqrt{\frac{\varepsilon}{8\pi}} \int_{\psi_s}^{\psi(z)}\!\!\!\!\!\!\!\!\frac{{\rm d}\psi}{\sqrt{P_0(\psi) -P_0(\psi{=}0)}},
\end{equation}
where $P_0$ is given by Eq.~(\ref{bcskygk2}), and we assumed that the bulk solution corresponds to the vanishing electrostatic potential, $\psi=0$.
In terms of rescaled (dimensionless) parameters, the above integral yields $z/\lambda_{\rm D}$ as a function of the reduced potential, $\beta e \psi$, with  $\zeta, \gamma$, $z_p$ and $\Xi = \lambda_{\rm D} /\ell_{\rm GC}$ as adjustable parameters.

The spatial profiles of the ionic densities, $en_{+}(z), -en_{-}(z)$ and $e_p(z)\, p(z)$ are plotted in Fig.~\ref{Fig4}. These densities illustrate how the screening of external fixed surface charge is partitioned between the monovalent salt and the dissociable macro-ions. For $\zeta \to 0$ (no macro-ions), one recovers the standard double-layer description, but as $\zeta$ increases, the most substantial contribution to the screening of the surface charge is progressively undertaken by the macro-ions. This leads to a much less pronounced spatial variation in the monovalent salt concentration, until most of the surface charge is actually screened by the dissociated cations and the remaining negatively charged macro-ions.

\begin{figure}[t!]
\centerline{\includegraphics[width=.3\textwidth]{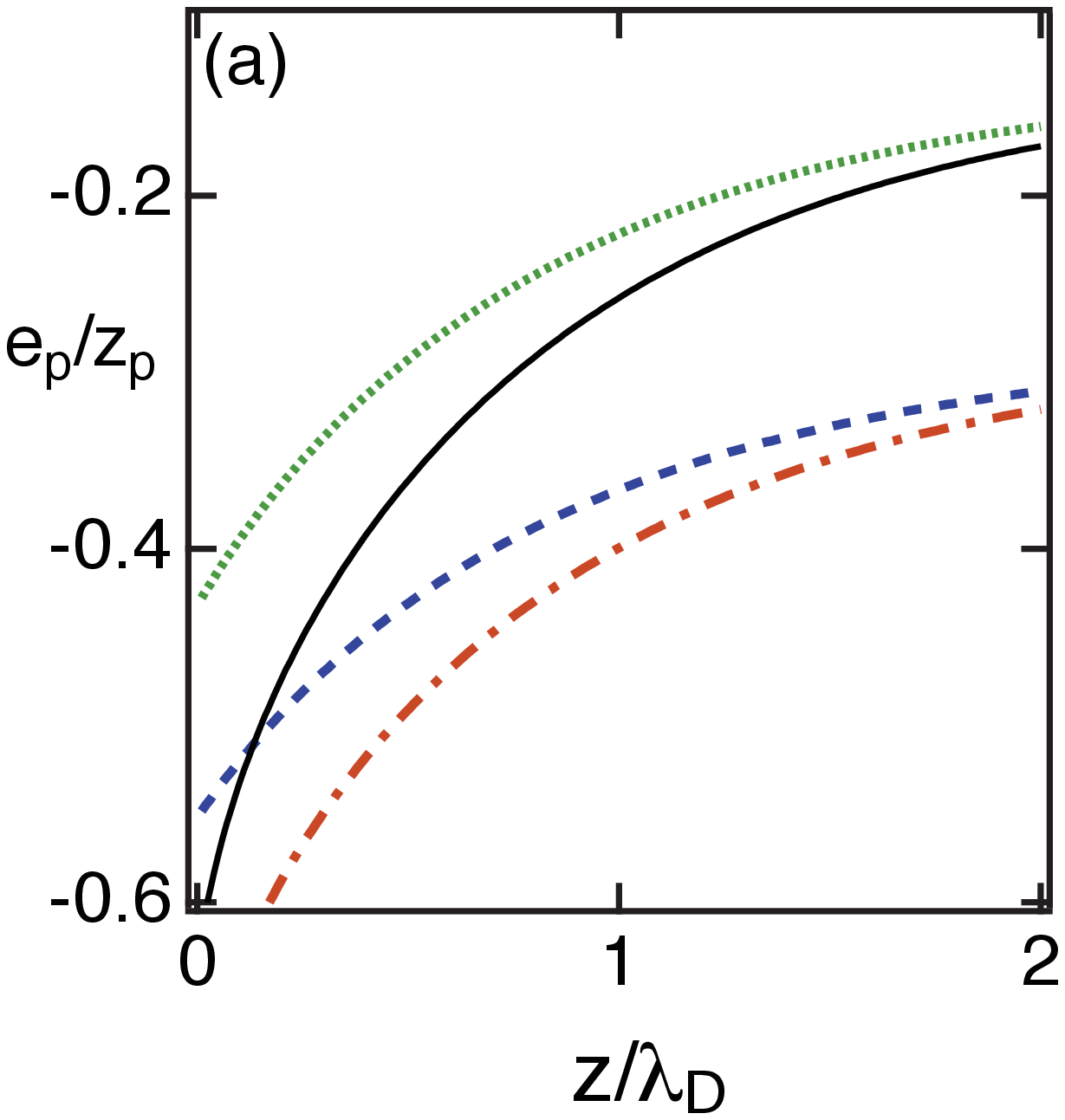}}
\centerline{\includegraphics[width=.3\textwidth]{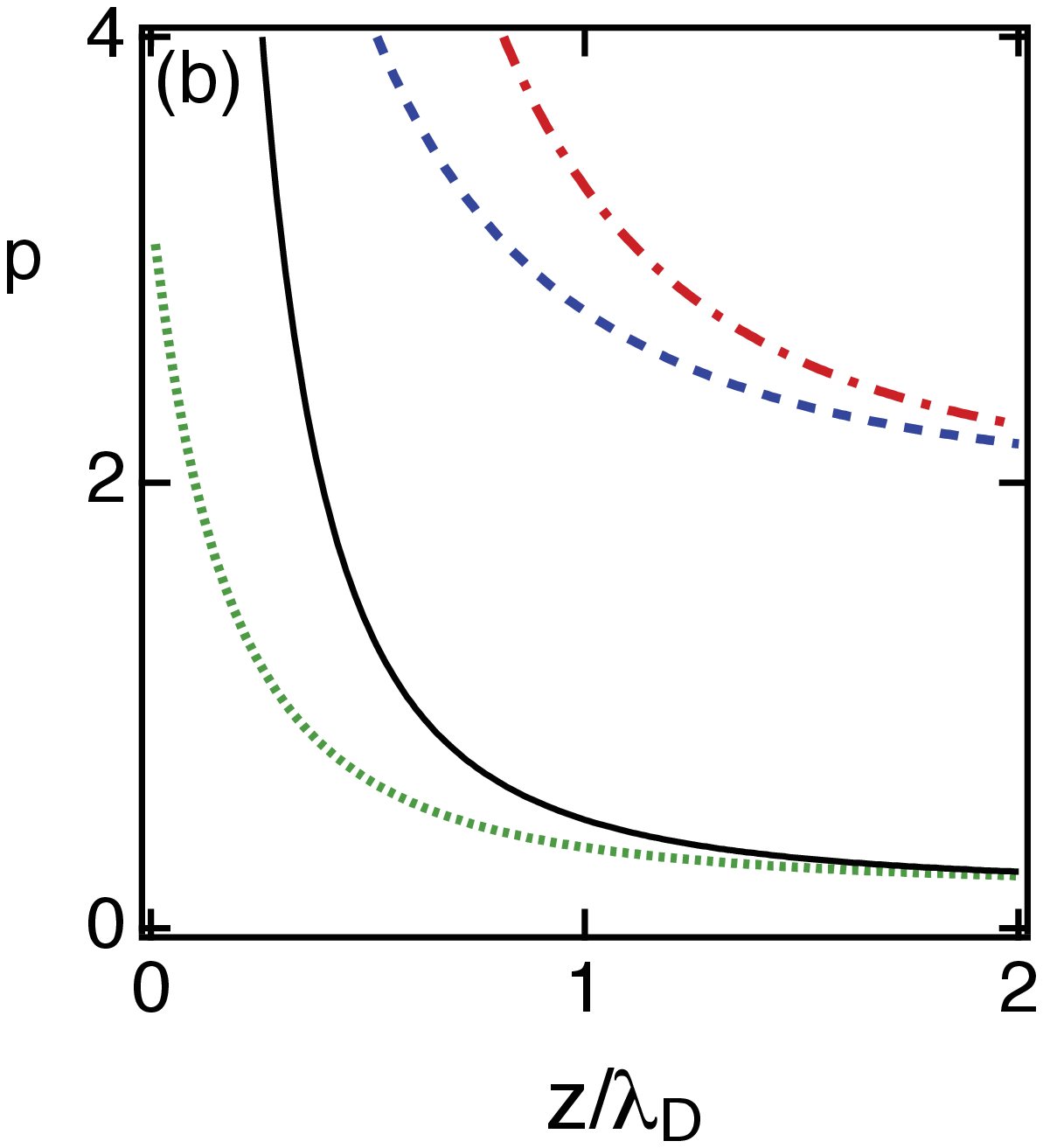}}
\caption{\textsf{(a) The macro-ion effective charge, $e_p = {\partial g}/{\partial \psi}$, in units of $z_p$, as function of the rescaled distance $ z/\lambda_{\rm D}$. (b) The macro-ion number concentration $p$, as function of the rescaled distance, $z/\lambda_{\rm D}$. The used parameters are: $\zeta = 2$, $\Xi=1$ (dotted green line), $\zeta = 2$, $\Xi=5$ (full black line), $\zeta = 4$, $\Xi=1$ (dashed blue line), and $\zeta = 4$, $\Xi=5$ (dash-dotted red line), and $\gamma = 0.5$, $z_p = 10$.}
\label{Fig5}}
\end{figure}

We note that even if the dissociation of the macro-ions in the bulk is not complete, the macro-ions release more cations into the solution when they are in the vicinity of the charged surface. This is clearly seen in Fig.~\ref{Fig5} from the dependence of the macro-ion effective charge, $e_p$, on the spatial position. Within the electric double layer, as the macro-ions approach the charged surface, the dissociation fraction will increase (Fig.~\ref{Fig5}\,a) together with an increase in the macro-ion concentration (Fig.~\ref{Fig5}\,b).

The structure of the double layer with dissociable macro-ions only partially depends on the redistribution of the salt ions. Near the surface but beyond the vicinal region of thickness larger than the macro-ion size, the double layer is progressively governed by the local dissociation equilibrium of the macro-ions.
At increasingly larger distances from the charged surface, the mean potential is small enough so that the linearization of the PB equation is permissible. Then, the mobile charge distribution is characterized by the effective screening length, $\lambda_{\rm eff}$, as discussed above. Hence, the screening of external fixed charges in presence of dissociable macro-ions is an adjustable variable, in contrast to the standard PB model.
This screening depends both on the monovalent salt ions of fixed valency as well as on the {\it chargeable} macro-ions of a variable valency.

\section{Conclusions}
The properties of charged regulating macro-ions,  as introduced in the present work, are of great relevance to a variety of complex fluids and colloidal solutions, where the macro-ion component contains dissociable moieties. For example,  solutions of proteins, with dissociable amino acids, or charged nano-particles with dissociable surface groups, can exhibit features that are {\it not shared} by more common ionic solutions, where all ionic species  have fixed valencies.

In this Letter, we suggest a new framework to treat mobile macro-ions in an electrolyte solution.
With the use of an augmented free-energy that generalizes the commonly used PB free-energy in a simple, although conceptual,
manner we were able to obtain quite distinguished features even within the MF approximation.
The self-regulation of the macro-ions effective charge changes the charge distribution close to charged macro-molecules,
broadly exhibits non-monotonic screening length and may serve as a buffer for the small ions in the solution.
This features plays an important role towards a consistent rationalization and prediction of various assembly processes.

Our novel results were obtained using a simplified charge regulation model.  We believe that the use of our framework for more elaborated charge regulation processes and other refinements will shed new light on the properties of complex solutions consisting of both CR macro-ions and simple salts. For example, it would be useful to investigate the charge regulation of mobile macro-ions consisting of several separate dissociable sites or even sites that compete for charge association/dissociation reactions. Such processes can exhibit many unexpected features in the charging equilibrium and require further elucidation.

\acknowledgments
We thank R. Adar and Y. Avni for useful discussions and numerous suggestions. This work was supported in part by the Israel Science Foundation (ISF) under Grant No. 438/12, the US-Israel Binational Science Foundation (BSF) under Grant No. 2012/060, and the ISF-NSFC (China) joint research program under Grant No. 885/15. RP would like to thank the School of Physics \& Astronomy at Tel Aviv University for its hospitality, and is grateful for the Sackler Scholar award and the Nirit and Michael Shaoul fellowship, within the framework of the Mortimer and Raymond Sackler Institute of Advanced Studies at Tel Aviv University. TM acknowledges the support from the
Blavatnik postdoctoral fellowship programme.



\end{document}